# ASTRA: A Python Package for Cross-Instrument Stellar and Telluric Template Construction

André M. Silva [1,2], J. P. Faria [3], Nuno C. Santos [1,2], Sérgio G. Sousa [1], Pedro T. P. Viana [1,2], and J. H. C. Martins [1]

**1** Instituto de Astrofísica e Ciências do Espaço, CAUP, Universidade do Porto, Rua das Estrelas, 4150-762 Porto, Portugal **2** Departamento de Física e Astronomia, Faculdade de Ciências, Universidade do Porto, Rua do Campo Alegre, 4169-007 Porto, Portugal **3** Département d'astronomie de l'Université de Genève, Chemin Pegasi 51, 1290 Versoix, Switzerland

## Summary

*ASTRA* is a Python package that provides a modular, instrument-independent interface for working with high-resolution stellar spectra. Designed to support data from multiple spectrographs—including ESPRESSO (Pepe et al., 2021), HARPS (Mayor et al., 2003; Pepe et al., 2002), MAROON-X (Seifahrt et al., 2022), and CARMENES (Quirrenbach et al., 2014)—*ASTRA* offers a unified abstraction over their data formats, enabling consistent access to fluxes, wavelengths, uncertainties, and metadata across instruments. Furthermore, it applies the necessary wavelength and flux calibrations that are needed, as described by the official pipelines of each instrument.

In addition to a common interface, *ASTRA* provides internal quality control checks of the observations, automatically divides them into the different sub-datasets that are commonly used by each spectrograph, and provides avenues to dynamically reject observations based on different properties. Furthermore, it also provides routines to mask the spectral imprint of Earth's atmosphere (in the form of telluric lines) and construct high-SNR, data-driven, stellar templates.

This package serves as the backend of the sBART pipeline (Silva et al., 2022) and is designed to be extensible and suitable for integration into larger spectral analysis workflows, enabling the construction of pipelines without having to tailor them to individual instruments. It is implemented in such a way that the user can select to only open in memory a small number of observations, such that it can seamlessly handle datasets with thousands of observations. Furthermore, it makes use of Python's autoproxy objects, ensuring a smooth integration with codes that use the *multiprocessing* library to leverage concurrent processing for faster computations.

## Statement of need

In recent years, multiple ultra-stable, high-resolution spectrographs capable of meter-per-second (or better) radial velocity precision have become central to exoplanet and stellar astrophysics, e.g. HARPS, CARMENES, MAROON-X, and ESPRESSO. While each spectrograph provides high-quality observations, they also use distinct data formats and apply different corrections to the stellar spectra. This hinders the development of generalized analysis pipelines, and often leads to analysis pipelines that are focused on a single instrument. To the best of our knowledge, there is no library that provides a similar interface to access stellar spectra of multiple state-of-the-art spectrographs.

---

*ASTRA* was built with the intention of not only providing a common API to access stellar spectra, but also provides:

1. Management of observations – *ASTRA* provides a high-level interface for accessing stellar spectra and metadata, built on a memory-efficient design that only loads a minimal number of spectra at a time. This ensures that *ASTRA* stays responsive, even when dealing with datasets with thousands of observations, at the cost of computational speed when interfacing with the data. This is done through the *autoproxy* interface, ensuring full compatibility with any *multiprocessing*-based application.
2. Masking of atmospheric features – When dealing with ground-based spectroscopy, it is important to account for the spectral imprint of our atmosphere (in the form of telluric lines), as well as its yearly variation. *ASTRA* can automatically run *TelFit* (Gullikson et al., 2014) to generate a synthetic transmittance model and create a binary mask to reject the position of telluric lines;
3. Construction of high-SNR stellar models – The construction of high-SNR stellar templates from observations is pivotal for the extraction of precise radial velocities (Artigau et al., 2022; Silva et al., 2022; Zechmeister et al., 2018), determination of stellar parameters (Gomes da Silva, J. et al., 2021; Sousa, S. G. et al., 2024), and characterization of exoplanetary atmospheres (Azevedo Silva et al., 2022; Damasceno, Y. C. et al., 2024; Stangret, M. et al., 2024);
4. Dynamically group the observations into different sub-datasets – Over their lifetime, most state-of-the-art spectrographs are subjected to instrumental interventions, leading to changes in the instrumental profile and offsets in radial velocities. As a consequence, it is often necessary to divide our data in the time-periods before and after such interventions, to construct individual templates in each. *ASTRA* is pre-configured with the dates of such instruments for all supported spectrographs, divides the observations in each dataset (or sub-instrument) and creates individual stellar and telluric templates for each;
5. Filtering of observations – When analysing data, we often reject observations based on metadata information (weather conditions, airmass, among others). Within *ASTRA* the user can dynamically set filters on different properties, with the goal of either fully rejecting the observation, or rejecting it for a specific operation. This means that it is possible to reject an observation for the construction of the stellar template, but not reject it from any subsequent analysis.
6. Masking of wavelength regions – When dealing with stellar spectra we often need to reject wavelength regions due to different contaminating effects. *ASTRA* creates an internal binary mask for all pixels and allows the rejection of i) telluric-contaminated regions; ii) activity-sensitive regions; iii) user-provided wavelength intervals.

As the backend of the sBART pipeline, it is already in use for scientific production, and is well-positioned to support the broader astrophysical community working with high-resolution spectroscopy.

# Acknowledgements

This work was funded by the European Union (ERC, FIERCE, 101052347). Views and opinions expressed are however those of the author(s) only and do not necessarily reflect those of the European Union or the European Research Council. Neither the European Union nor the granting authority can be held responsible for them. This work was also supported by FCT - Fundação para a Ciência e a Tecnologia through national funds by these grants: UIDB/04434/2020 DOI: 10.54499/UIDB/04434/2020, UIDP/04434/2020 DOI: 10.54499/UIDP/04434/2020, PTDC/FIS-AST/4862/2020, UID/04434/2025. JPF acknowledges the financial support of the Swiss National Science Foundation (SNSF), supported since May 2022 over the grant 200020_20501.

# References

Artigau, E., Cadieux, C., Cook, N. J., Doyon, R., Vandal, T., Donati, J.-F., Moutou, C., Delfosse, X., Fouqué, P., Martioli, E., Bouchy, F., Parsons, J., Carmona, A., Dumusque, X., Astudillo-Defru, N., Bonfils, X., & Mignon, L. (2022). Line-by-line velocity measurements: An outlier-resistant method for precision velocimetry. *The Astronomical Journal*, *164*(3), 84. https://doi.org/10.3847/1538-3881/ac7ce6

Azevedo Silva, T., Demangeon, O. D. S., Santos, N. C., Allart, R., Borsa, F., Cristo, E., Esparza-Borges, E., Seidel, J. V., Palle, E., Sousa, S. G., Tabernero, H. M., Zapatero Osorio, M. R., Cristiani, S., Pepe, F., Rebolo, R., Adibekyan, V., Alibert, Y., Barros, S. C. C., Bouchy, F., … Udry, S. (2022). Detection of barium in the atmospheres of the ultra-hot gas giants WASP-76b and WASP-121b: Together with new detections of Co and Sr+ on WASP-121b. *Astronomy & Astrophysics*, *666*, L10. https://doi.org/10.1051/0004-6361/202244489

Damasceno, Y. C., Seidel, J. V., Prinoth, B., Psaridi, A., Esparza-Borges, E., Stangret, M., Santos, N. C., Zapatero-Osorio, M. R., Alibert, Y., Allart, R., Azevedo Silva, T., Cointepas, M., Costa Silva, A. R., Cristo, E., Di Marcantonio, P., Ehrenreich, D., González Hernández, J. I., Herrero-Cisneros, E., Lendl, M., … Pepe, F. (2024). The atmospheric composition of the ultra-hot Jupiter WASP-178 b observed with ESPRESSO⋆. *Astronomy & Astrophysics*, *689*, A54. https://doi.org/10.1051/0004-6361/202450119

Gomes da Silva, J., Santos, N. C., Adibekyan, V., Sousa, S. G., Campante, T. L., Figueira, P., Bossini, D., Delgado-Mena, E., Monteiro, M. J. P. F. G., de Laverny, P., Recio-Blanco, A., & Lovis, C. (2021). Stellar chromospheric activity of 1674 FGK stars from the AMBRE-HARPS sample - i. A catalogue of homogeneous chromospheric activity⋆. *Astronomy & Astrophysics*, *646*, A77. https://doi.org/10.1051/0004-6361/202039765

Gullikson, K., Dodson-Robinson, S., & Kraus, A. (2014). Correcting for telluric absorption: Methods, case studies, and release of the TelFit code. *The Astronomical Journal*, *148*(3), 53. https://doi.org/10.1088/0004-6256/148/3/53

Mayor, M., Pepe, F., Queloz, D., Bouchy, F., Rupprecht, G., Lo Curto, G., Avila, G., Benz, W., Bertaux, J.-L., Bonfils, X., Dall, Th., Dekker, H., Delabre, B., Eckert, W., Fleury, M., Gilliotte, A., Gojak, D., Guzman, J. C., Kohler, D., … Weilenmann, U. (2003). Setting new standards with HARPS. *The Messenger*, *114*, 20–24.

Pepe, F., Cristiani, S., Rebolo, R., Santos, N. C., Dekker, H., Cabral, A., Di Marcantonio, P., Figueira, P., Curto, G. L., Lovis, C., Mayor, M., Mégevand, D., Molaro, P., Riva, M., Osorio, M. R. Z., Amate, M., Manescau, A., Pasquini, L., Zerbi, F. M., … Zanutta, A. (2021). ESPRESSO@VLT – On-sky performance and first results. *Astronomy & Astrophysics*, *645*, A96. https://doi.org/10.1051/0004-6361/202038306

Pepe, F., Mayor, M., Rupprecht, G., Avila, G., Ballester, P., Beckers, J.-L., Benz, W., Bertaux, J.-L., Bouchy, F., Buzzoni, B., Cavadore, C., Deiries, S., Dekker, H., Delabre, B., D'Odorico, S., Eckert, W., Fischer, J., Fleury, M., George, M., … Penny, A. (2002). HARPS: ESO's coming planet searcher. Chasing exoplanets with the La Silla 3.6-m telescope. *The Messenger*, *110*. https://ui.adsabs.harvard.edu/abs/2002Msngr.110....9P

Quirrenbach, A., Amado, P. J., Caballero, J. A., Mundt, R., Reiners, A., Ribas, I., Seifert, W., Abril, M., Aceituno, J., Alonso-Floriano, F. J., Eiff, M. A., Jiménez, R. A., Anwand-Heerwart, H., Azzaro, M., Bauer, F., Barrado, D., Becerril, S., Béjar, V. J. S., Benítez, D., … Xu, W. (2014). CARMENES instrument overview. In S. K. Ramsay, I. S. McLean, & H. Takami (Eds.), *Ground-based and airborne instrumentation for astronomy v* (Vol. 9147, p. 91471F). International Society for Optics; Photonics; SPIE. https://doi.org/10.1117/12.2056453

Seifahrt, A., Bean, J. L., Kasper, D., Stürmer, J., Brady, M., Liu, R., Zechmeister, M., Stefánsson, G. K., Montet, B., White, J., Tapia, E., Mocnik, T., Xu, S., & Schwab, C. (2022). MAROON-X: The first two years of EPRVs from Gemini North. In C. J.

Evans, J. J. Bryant, & K. Motohara (Eds.), *Ground-based and airborne instrumentation for astronomy IX* (Vol. 12184, p. 121841G). International Society for Optics; Photonics; SPIE. https://doi.org/10.1117/12.2629428

Silva, A. M., Faria, J. P., Santos, N. C., Sousa, S. G., Viana, P. T. P., Martins, J. H. C., Figueira, P., Lovis, C., Pepe, F., Cristiani, S., Rebolo, R., Allart, R., Cabral, A., Mehner, A., Sozzetti, A., Mascareño, A. S., Martins, C. J. A. P., Ehrenreich, D., Mégevand, D., … al, et. (2022). A novel framework for semi-Bayesian radial velocities through template matching. *Astronomy & Astrophysics*. https://doi.org/10.1051/0004-6361/202142262

Sousa, S. G., Adibekyan, V., Delgado-Mena, E., Santos, N. C., Rojas-Ayala, B., Barros, S. C. C., Demangeon, O. D. S., Hoyer, S., Israelian, G., Mortier, A., Soares, B. M. T. B., & Tsantaki, M. (2024). SWEET-cat: A view on the planetary mass-radius relation. *Astronomy & Astrophysics*, *691*, A53. https://doi.org/10.1051/0004-6361/202451704

Stangret, M., Palle, E., Esparza-Borges, E., Orell Miquel, J., Casasayas-Barris, N., Zapatero Osorio, M. R., Cristo, E., Allart, R., Alibert, Y., Borsa, F., Demangeon, O. D. S., Di Marcantonio, P., Ehrenreich, D., Figueira, P., González Hernández, J. I., Herrero-Cisneros, E., Martins, C. J. A. P., Santos, N. C., Seidel, J. V., … Udry, S. (2024). The obliquity and atmosphere of the hot Jupiter WASP-122b (KELT-14b) with ESPRESSO: An aligned orbit and no sign of atomic or molecular absorption⋆. *Astronomy & Astrophysics*, *691*, A120. https://doi.org/10.1051/0004-6361/202450938

Zechmeister, M., Reiners, A., Amado, P. J., Azzaro, M., Bauer, F. F., Béjar, V. J. S., Caballero, J. A., Guenther, E. W., Hagen, H.-J., Jeffers, S. V., Kaminski, A., Kürster, M., Launhardt, R., Montes, D., Morales, J. C., Quirrenbach, A., Reffert, S., Ribas, I., Seifert, W., … Wolthoff, V. (2018). Spectrum radial velocity analyser (SERVAL): High-precision radial velocities and two alternative spectral indicators. *Astronomy & Astrophysics*, *609*, A12. https://doi.org/10.1051/0004-6361/201731483